\documentclass[conference]{IEEEtran}
\usepackage{amsmath}
\usepackage[usenames,dvipsnames,svgnames,table]{xcolor}
\usepackage{url}
\usepackage[pdftex]{graphicx}
\usepackage[utf8]{inputenc}
\usepackage{mathptmx}
\usepackage{tabularx}
\usepackage{ragged2e}
\usepackage[singlelinecheck=false]{caption}
\usepackage[T1]{fontenc}
\usepackage[numbers]{natbib}
\usepackage{amsmath}
\usepackage{booktabs}
\usepackage{colortbl}
\usepackage{float}
\usepackage{flushend}
\usepackage{soul}
\usepackage{subcaption}
\usepackage{siunitx}
\usepackage{listings}
\usepackage{tcolorbox}
\tcbuselibrary{xparse}

\begin{document}
\DeclareTotalTCBox{\myverb}{ v }{verbatim, colframe=blue!75!black,colupper=blue}{#1}
\title{Transparent Live Code Offloading on FPGA}

\author{\IEEEauthorblockN{Roberto Rigamonti and Baptiste Delporte and Anthony Convers and Alberto Dassatti}
\IEEEauthorblockA{HES-SO\\REDS Institute, HEIG-VD --- School of Business and Engineering Vaud\\
  CH-1400 Yverdon-les-Bains, Switzerland (\emph{name.surname@heig-vd.ch})}}

\maketitle

\begin{abstract}
  Even though it seems that FPGAs have finally made the transition from research labs to the consumer devices' market,
  programming them remains challenging.
  Despite the improvements made by High-Level Synthesis (HLS), which removed the language and paradigm barriers that
  prevented many computer scientists from working with them, producing a new design typically requires at least several hours,
  making data- and context-dependent adaptations virtually impossible.

  In this paper we present a new framework that off-loads, on-the-fly and transparently to
  both the user and the developer, computationally-intensive code fragments to FPGAs.
  While the performance should not surpass that of hand-crafted HDL code,
  or even code produced by HLS, our results come with no additional development costs
  and do not require producing and deploying a new bit-stream to the FPGA each time a
  change is made.
  Moreover, since optimizations are made at run-time, they may fit particular datasets or
  usage scenarios, something which is rarely foreseeable at design or compile time.

  Our proposal revolves around an overlay architecture that is pre-programmed on the FPGA and
  dynamically reconfigured by our framework to execute code fragments extracted from the
  Data Flow Graph (DFG) of computational intensive routines.
  We validated our solution using standard benchmarks and proved we are able to off-load
  to FPGAs without developer's intervention.
\end{abstract}

\IEEEpeerreviewmaketitle

\section{Introduction}
\noindent Heterogeneous computing has recently emerged as a way to circumvent the physical and technological limitations in the design of computing devices~\cite{Brodtkorb10}.
The pressure exerted from the growing demand for better performance has finally made the long-awaited dream of having a traditional CPU paired with an FPGA a
reality~\cite{Crockett14,Alam07}.
Using FPGAs has indeed been proven to be viable for increasing energy efficiency in High-Performance Computing (HPC)~\cite{Putnam14,Canis11,Chen16}.
However, while providing the systems integrator with a compact and cost effective way to add advanced functionalities to products, this technological evolution has dramatically raised the
overall complexity of systems:
Exploiting new capabilities now requires a wide range of competencies that are rarely possessed by companies and institutions that need them.
Due to the considerable effort demanded at development time, applicability is often limited to a reduced set of supported brands/models,
while being effective only when the predicted usage patterns match the actual ones.
HLS~\cite{Gajski12,Martin09} partially mitigates these problems by removing the language barrier. However, compiling and deploying a bit-stream is an extremely
long process, and HLS development requires the establishment of a-priori usage patterns that might lead to  sub-optimal usage of available hardware.

We propose an automated framework that allows the transparent execution of ordinary code on a heterogeneous platform including an FPGA.
Our solution requires no changes to application code, not even \texttt{pragma} indications to guide the optimization (although we can benefit from their presence),
and dynamically adapts its behavior to the execution scenario and workload of the system.
Therefore, our approach relieves the developer from the burden of being aware of the target platform's details.
Moreover, she does not have to forecast use cases to prevent performance bottlenecks, nor does she have to statically decide
which parts of the system have to be accelerated: The system transparently
identifies parallelizable, computationally-intensive code fragments and dispatches them to a data-flow overlay engine pre-programmed on the FPGA.
Since the bit-stream we use is fixed, and in contrast with HLS, we can alter the functionalities offered by the FPGA on-the-fly, to adapt them to current usage patterns.
Finally, since we operate at the LLVM's Intermediate Representation (IR) level~\cite{Lattner11},
our approach is language-agnostic.

\begin{figure}[t]
  \centering
  \includegraphics[width=\columnwidth]{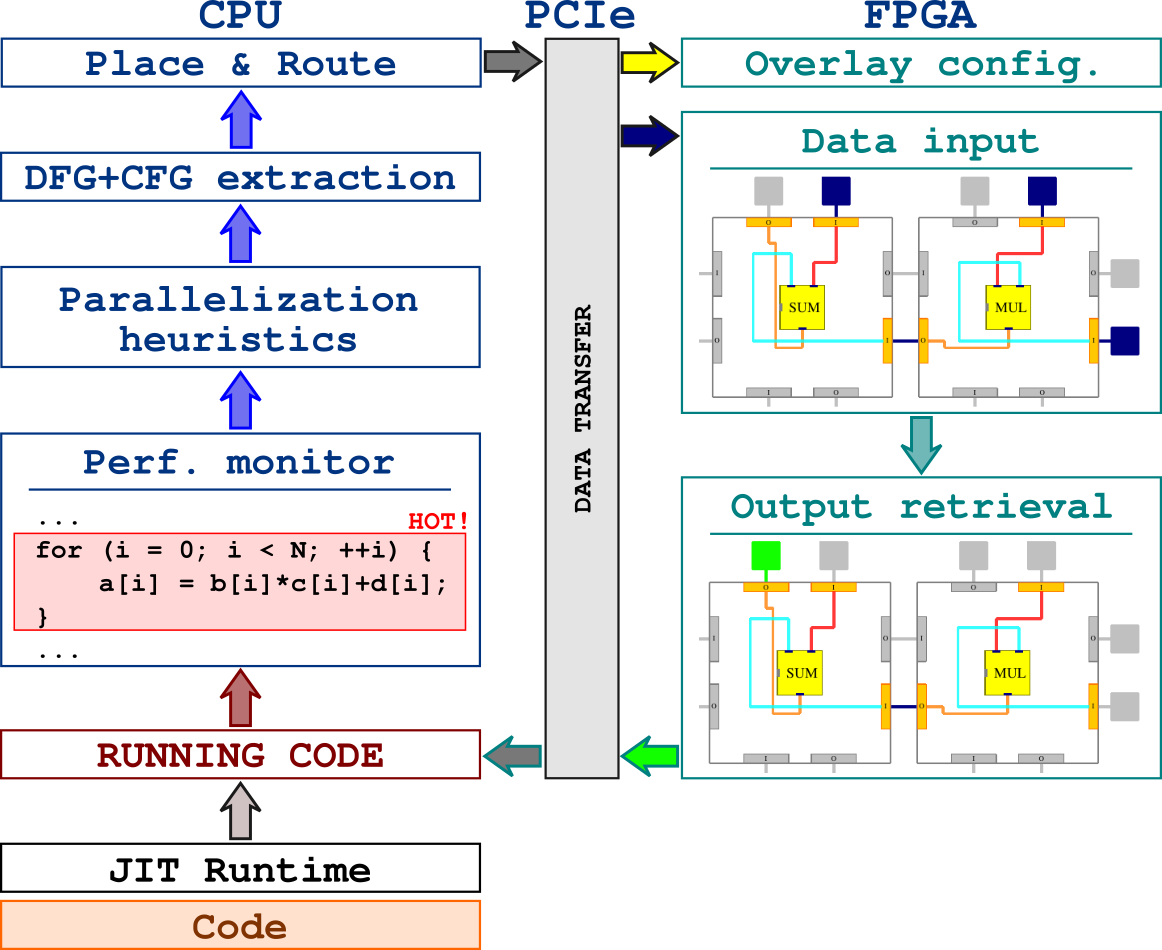}
  \caption{Schematic representation of the developed system.
    The code in execution is monitored to detect computationally-intensive fragments. If these fragments are susceptible of optimization, the overlay programmed on the FPGA (DFE) is reconfigured and
    the required data is sent. Once computed, the results are fed back to the running code.}
  \label{fig:tfa}
\end{figure}

At the heart of our system, depicted by Fig.~\ref{fig:tfa}, lies a Just-In-Time (JIT) compiler, coupled with a low-overhead performance monitor to
automatically detect which code fragments require the largest fraction of resources (namely execution time or memory accesses).
The usage of a JIT framework is key to our approach.
In addition to being able to perform optimizations that can only be applied at run-time, one can detect which parts of the code are actually taking the largest fraction of resources
based on current inputs.
With this information we can avoid off-loading unimportant code fragments that would result in little gain, while still retaining the power to revert these decisions
should these fragments acquire more relevance.

Once a code region is identified as critical, it is analyzed to expose parallelization opportunities.
The Control Flow Graph (CFG) and the Data Flow Graph (DFG) are then extracted and used to drive the placement and routing of functional units over the overlay,
which will be called Data Flow Engine (DFE) in the following.
Finally, the FPGA's overlay is reconfigured on-the-fly to execute the new data flow model.
Once this is done (it requires few milliseconds), we alter the execution flow of the code as in~\cite{Delporte15a} and feed to the FPGA
the data provided by the running application.

An important characteristic of our proposal is that we are not tied to any specific hardware solution. We support all FPGAs, and the DFE we adopted has a parametric size
to adapt to the available resources of different devices.  Moreover, the chosen overlay itself is just a choice of convenience, as our framework architecture is generic.
For this reason we have invested low effort in optimizing it, even though multiple easy optimization opportunities are clearly available --- for example, blocks could be specialized to perform
only specific operations, or the design could be manually placed~\cite{Capalija13}.

Our long-term vision is to develop a framework that significantly reduces development time by allowing the code to be written just once, in a form most natural for the high-level developer,
and then optimized on-the-fly only when needed and according to available hardware capabilities.
This would be a considerable improvement over current HLS techniques, and the work presented in this paper represents the first step in this direction.
To the best of our knowledge, no other approach proposed thus far is capable of achieving these objectives.

\section{Related Work}
The increasing heterogeneity of computing devices has originated a prolific research community whose ultimate goal is to transpose hardware improvements in actual speed-ups for the final user.
The vast majority of proposed approaches rely heavily on the high-level software designers' skills in mapping hardware capabilities to the software fragments where they are most needed.

One of the earlier examples is given by OpenMP~\cite{Dagum98}, a directive-based API for shared-memory multiprocessor programming. While OpenMP is the de-facto standard for scientific computing on these platforms, it only recently (from version 4.0 with Heterogeneous Parallel Programming extensions) approached   heterogeneous platforms. Due to this late response, some alternatives emerged, among which
OpenACC~\cite{Wienke12} is the most recent and most promising one.
Initiated by some of the major players in the HPC arena, it allows C/C++ and Fortran code fragments to be off-loaded to external accelerators while the main code runs on the host.
While relieved of a large part of the language burden imposed by lower-level APIs, the developer is still required to define use cases to identify potential performance bottlenecks,
respect data dependencies, and make hypotheses on the expected workload.
Also, its performance has been reported to vary significantly with respect to the corresponding best-effort OpenCL implementation, scoring worse when the task to solve becomes more complex~\cite{Wienke12}.
OmpSs~\cite{Bueno12} pushes the level of parallelism achievable with a directive-based paradigm even further, allowing a fragment of code to be off-loaded to clusters of GPUs, but
suffers from the same issues mentioned above.

Restricting our focus to more specific application types,
a prominent role is played by Domain-Specific Languages (DSLs)~\cite{Fowler10}.
Their purpose is to map high-level code to heterogeneous systems, allowing the developer to work with an implicitly parallel language.
Delite~\cite{Chafi11}, for instance, is a framework for the development of DSLs that includes a dynamic run-time capable of executing the developed code on a CPU+GPU system.
It allows writing a DSL for a specific domain --- in~\cite{Fowler10} an example DSL for machine learning applications, called OptiML, is presented too ---, thus representing a good alternative to
directive-based approaches.
However, it requires high-level software developers to be proficient in an additional language (to which they have most likely not been exposed beforehand, since it
has been designed for the particular application domain).

More specialized use cases still rely upon low-level API programming, with CUDA~\cite{Nickolls08} and OpenCL~\cite{Stone10} being the most renowned ones.
The former is tightly coupled to NVIDIA GPUs, the latter is nowadays gaining popularity as a high-level interface towards FPGAs~\cite{Settle13} and GPUs.
Using OpenCL to program FPGA represents a big improvement over proprietary High-Level Synthesis (HLS) tools previously adopted, since OpenCL is an open and royalty-free standard that allows
developers to migrate, with little effort, their code from GPUs to FPGAs.
Nevertheless, OpenCL coding requires skills and a deep understanding of the memory-model behind it, and forces the developers to identify potential computationally-intensive code fragments.
Also, being a development-time approach, by its nature it cannot adapt to the input data and instantaneous computational load.

Relieving developers from code annotation and restructuring requires
automatic extraction of parallelism, a very complex task in which considerable efforts have been invested over the last decades~\cite{Banerjee93}.
Solutions have been proposed in two main branches: speculative approaches and analytical ones.
The former category is the most attractive, as it has been shown that analytical investigations into potential parallelization opportunities miss a large fraction of them~\cite{Aldea16}.
ATLaS~\cite{Aldea16} is an example of thread-level speculation, where OpenMP has been augmented with an additional clause to mark speculative executions.
Despite a nice theoretical framework, however, speculative approaches have encountered little success in practical implementations.

Analytical solutions, which detect optimization opportunities by static analysis of code, have in Polly~\cite{Grosser12} one of its most successful implementations.
Polly is a polyhedral optimizer for automatic parallelization that operates at the IR level and starts by converting the code to a polyhedral representation~\cite{Bondhugula08}, in which it
detects Static Control Parts (SCoPs)~\cite{Damschen15a}.
It then identifies the subset of SCoPs matching a specific canonical form, and automatically generates via an LLVM pass SIMD and OpenMP code.
A recent, very interesting approach strongly linked with Polly, called Polly-ACC~\cite{Grosser16}, shares our goal of transparently making heterogeneous hardware capabilities available
to software that would otherwise not benefit from them, mostly for financial of software engineering reasons.
It proposes a new compiler that interacts with a standard LLVM-based compiler to generate multi-device binaries, automatically handling data management.
For the time being, only CUDA code generation is supported, but OpenCL implementation is ongoing.
Our proposed system has two main differences with respect to~\cite{Grosser16}:
We target FPGAs as accelerators and we perform optimizations at run-time, thus exploiting the wealth of information available at that stage.

Another technique aimed at relieving the developer from the burden of adapting the code to heterogeneous environments is HPA~\cite{Delporte16a}, which is an energy-efficient
run-time optimizer that dynamically and transparently off-loads code fragments previously identified as worthy by a profiler.
The off-loading is performed by first transmitting the code (marked as parallelizable by Polly~\cite{Grosser12}) and the data
(identified by the use of custom allocators automatically put in place by the run-time) to the accelerator, and then altering the
main task --- which is running in a JIT framework --- to transfer the execution.
Our proposal is in the same spirit: We also use a LLVM-JIT framework, we monitor the execution to identify hotspots, and we perform our optimization by off-loading the
detected code fragments to an accelerator, reverting our choice if the resulting performance is deemed insufficient. However, in this paper we target a different kind of accelerator (FPGA) with
very specific features and limitations.

Recent research~\cite{Inggs16,Putnam14} demonstrates the profitability of accelerating
computations exploiting FPGAs. Despite this, FPGA accelerators are still rare and market penetration is limited by a few factors, among which design costs and portability are the most remarkable.
\cite{Stitt2011} proposes the idea of an Intermediate Fabrics, a virtualization of FPGA resources reducing design costs and augmenting significantly design portability. Exploiting this innovative
solution still demands skills beyond those traditionally possessed by software engineers.

An approach bearing resemblance to ours is presented in~\cite{Vahid08}, where decompiled code fragments extracted from a running binary are dinamically translated into an FPGA's bit-stream
by dedicated CAD tools and transparently accelerate execution.
However, this approach is limited to very simple applications~\cite{Happe13}, and requires specific hardware (such as an on-chip CAD module).
Also, the CPU and the FPGA are not allowed to run concurrently (to prevent data coherence and consistency issues)~\cite{Bispo12}, severely affecting the parallelism achievable by the system.

A different approach to solving the productivity issue is in providing automatic tools able to translate from an high level language to Hardware Description Languages (HDLs).
Examples of projects proposing solutions along this path are HLS suites, such as LegUP~\cite{Canis11}, and commercial products like Catapult~\cite{catapult} and Xilinx HLS. While these tools
claim to reduce the skill set needed by a designer to exploit an FPGA, the final product is an HDL description of a circuit. Passing from this description to a working system, requires
specialized tools and competencies, and even when they are available, tool running times are on the order of hours, making HLS adoption still complex and limited.

Due to the practical limitations of HLS systems, many researchers approached the topic of designing a programmable layer on top of the FPGA fabric. This layer is often referenced as overlay or
Coarse Grain Reconfigurable Architectures (CGRAs)~\cite{Tessier2015}. The idea is programming the fine grain FPGA fabric once with a programmable design that can then be customized at run time to
execute different functionalities. This approach has the benefit of reducing the complexity of mapping an application on an accelerator and at the same time virtualizes, comparably
to~\cite{Stitt2011}, the FPGA resources, making the design easily portable to devices from different vendors. In spite of these advantages, all proposed overlays suffer from consuming more
resources when compared with manual HDL designs or HLS based solutions.

We decided to develop a custom overlay (DFE) and the related software utilities in order to develop a complete prototype of the proposed framework.

\section{Proposed Approach}
\begin{figure*}[t]
  \centering
  \includegraphics[width=\linewidth]{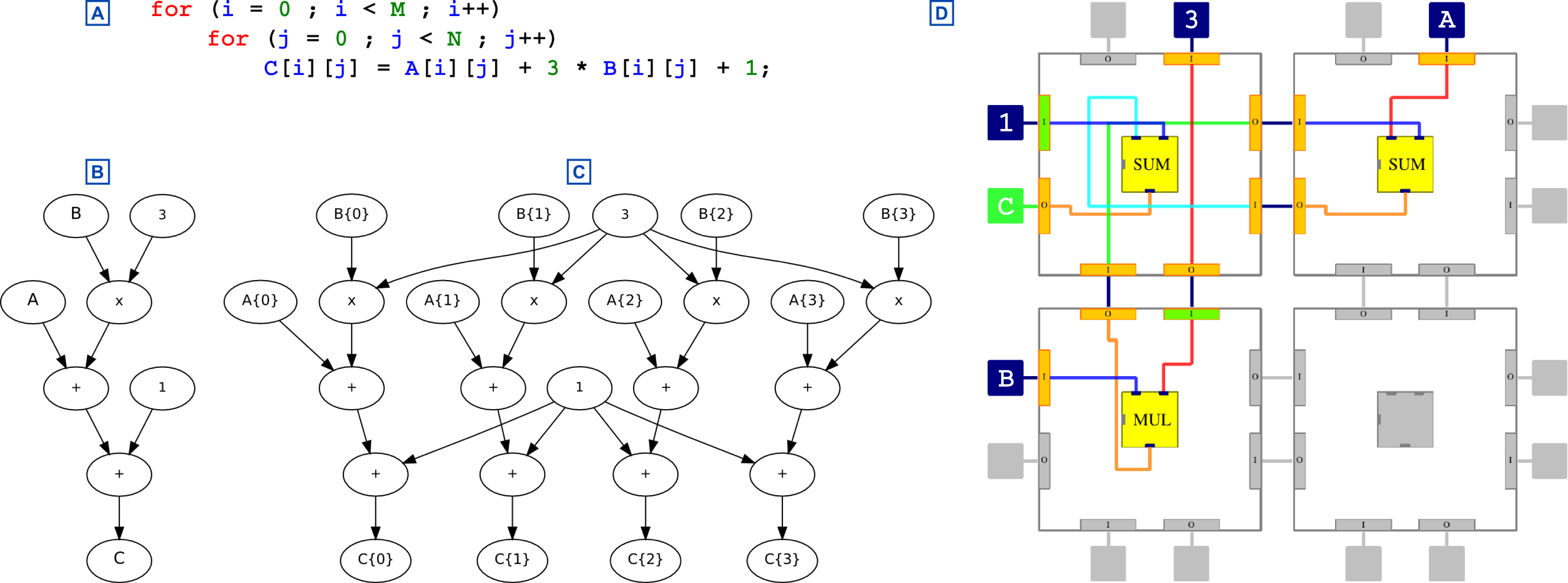}
  \caption{Example of transformation of a parallelizable C code into an overlay's configuration.
    \textbf{(A)} Sample code fragment to optimize, performing the matrix operation $C=A+3B+1$.
    \textbf{(B)} Corresponding DFG.
    \textbf{(C)} DFG when loop unrolling (here by a factor four) is applied. The numbers between braces are used to partition the input data.
    \textbf{(D)} Overlay configuration resulting from the placing\&routing of the computed DFG. Rectangular input boxes filled in green indicate constant
    values (in the example, ``1'' and ``3'') that are retained throughout the computations, reducing the amount of data to transfer.}
  \label{fig:example_pr}
\end{figure*}

In this section we detail the different steps undertaken by our system in order to transparently off-load part of an application in execution to an FPGA.

Our system takes as input the LLVM Intermediate Representation (IR) of an application.
Using LLVM-IR provides great flexibility to our system, in fact it becomes able to handle
applications written in any language for which a LLVM front-end exists (see~\cite{llvmwiki} for a complete list).
We could derive this input IR code from the binary code of an application using a decompilation framework such as McSema~\cite{Dinaburg14}.
Research in this direction is undergoing.
As illustrated by Fig.~\ref{fig:tfa}, the input code is translated by a JIT compiler into executable code for the host machine.
This phase could also be achieved without using a JIT compiler, but the system would have to collect detailed information about all functions' code and addresses.

Once the execution is ready, it is started under the supervision of a performance monitor that allows us to trace and detect hot-spots at run time.
Similarly to~\cite{Delporte16a}, we have chosen to use perf\_event~\cite{Weaver13}, which collects accurate statistics from both software and hardware counters and is well supported by the Linux kernel.
Based on simple metrics, such as computation time and memory accesses, the profiling sub-module selects interesting functions for the subsequent analysis phase.

Analysis starts by assessing if the identified code meets a series of criteria; For instance,
the system discards code requiring operations or data types not currently supported by any available hardware implementation.
Another possible criterion is the presence of system calls, as this would indicate that there are no optimization opportunities in a given code fragment.
If the code under test satisfies all these constraints, we move forward using a custom-made automatic parallelizer inspired by Polly to seek potential optimization opportunities.
If such an opportunity is found, we modify the IR view of the code accordingly.
At this stage, we extract and merge the CFG and DFG of the selected code and, if the number of nodes is larger than a certain threshold, we transfer the execution to the FPGA.
This latter decision is intended to discard small DFGs, for which it is highly probable that the data transfer overhead would negatively impact the overall performance. Threshold values must
be customized or experimentally determined for each implementation and, if the communication media is shared, could be dynamically modified in order to account for varying
system loads.
To transfer execution to an FPGA we need to start a potentially time-consuming task for placing and routing the functional units on our DFE.
This process is not deterministic and can require several seconds to complete.
Details about the chosen place\&route algorithm are given in Section~\ref{ss:pr}.
An example DFE configuration is given by Fig.~\ref{fig:example_pr}: \textbf{(A)} shows a fragment of code, in this particular case two nested for loops performing the matrix operation $C=A+3B+1$.
The corresponding DFG is depicted by \textbf{(B)}, while \textbf{(D)} shows the placed\&routed DFE (for graphical purposes we have selected a tiny $2 \times 2$ architecture).
Note that, according to the DFE's size and the considered problem, the DFG could easily be modified to support loop-unrolling and other standard optimizations,  as exemplified by \textbf{(C)}.

Once the DFE's configuration has been completed, the programming details are stored in a cache for later reuse.
We can indeed, on our prototype system, switch between different configurations in few milliseconds, so it makes sense to change configuration as often as needed.
Finally, the run-time replaces all calls to the host processor function with a wrapper stub that handles all memory transfers to and from the FPGA, and only then starts execution on it.

Instead of employing a sophisticated prediction model for estimating the performance of running a data flow code on the FPGA,
we continuously monitor the execution time and we roll back to the initial software should the produced implementation perform worse than the original one.
This approach guarantees complete adaptability to changing conditions of the system, while having a low overhead.

The above-mentioned steps are graphically presented in Fig.~\ref{fig:tfa}.
In particular, a path can be followed for a given code block from its input in the system to its execution on the DFE, and then until processed results are fed back into the running
application.

\subsection{DFE}
Up to this point, we have given a superficial overview of the implemented overlay.
As stated in the introduction, our approach is indeed independent and orthogonal with respect to the chosen overlay implementation.
In our framework, the chosen overlay is a plug-in and the only requirements we impose to an overlay in order to be exploitable are:
a) a series of constraints to be verified in order to quickly discard codes with specific undesirable features, and b)
a software place\&route routine able to take a DFG as input and produce the specific overlay configuration as output.
While we believe that customizing the overlay's architecture for specific applications is very promising, the implementation
choices we made thus far were directed at having a fully functional and measurable platform to prove our intuitions more than at having a specific
accelerator for one or a set of applications.

\begin{figure}[t]
  \centering
  \includegraphics[width=.6\columnwidth]{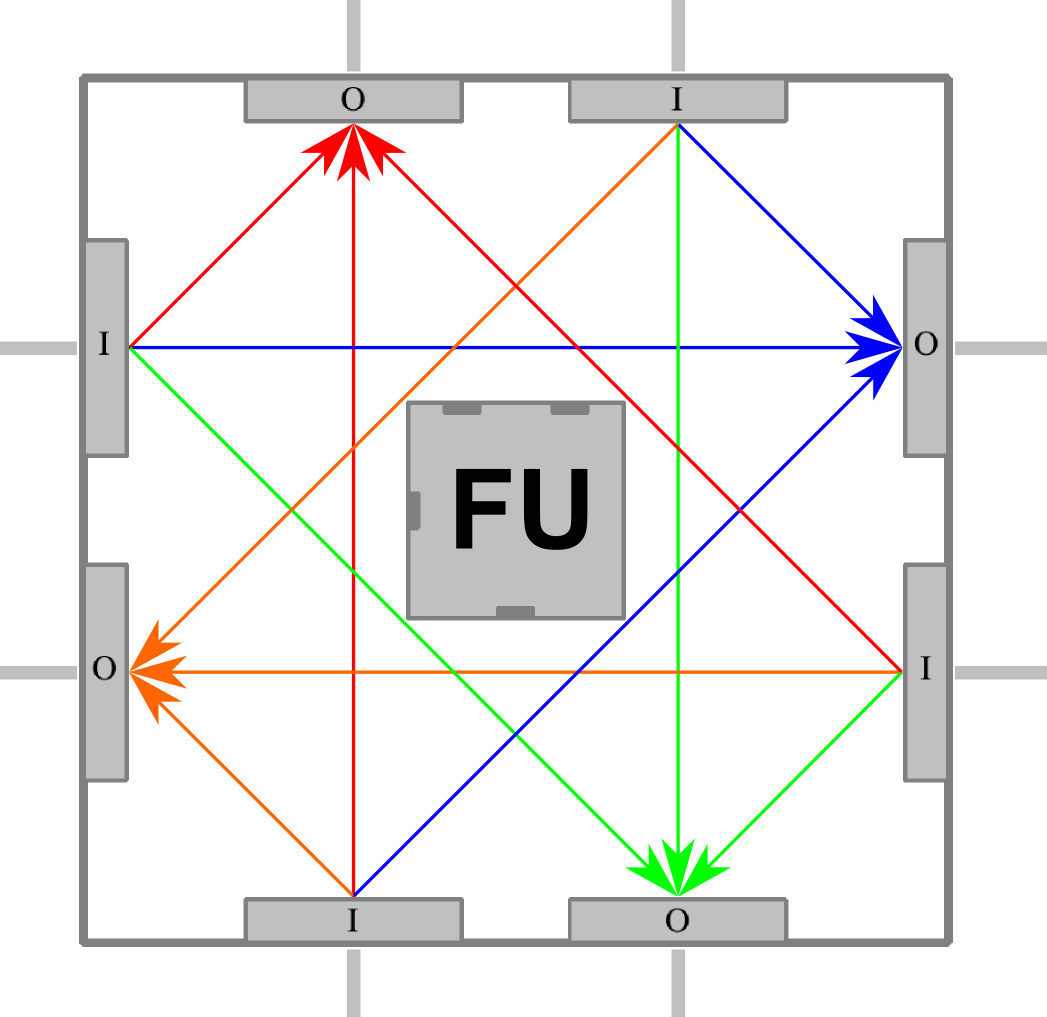}
  \caption{Cell's detailed view, with the possible direct connections between inputs and outputs depicted.
    Further connections (not shown) join each cell's input with the three Functional Unit's inputs (input 1, input 2, and selection input), and each cell's output with the Functional Unit's output.}
  \label{fig:pe}
\end{figure}

Our first prototype was developed in parametric VHDL in order to experiment with different FPGA chips and DFE sizes, and is based on the overlay presented in~\cite{Capalija13}.
This architecture involves a fully pipelined data flow overlay with a rich set of routing resources. Each node is a functional unit that can be
configured to execute an operation on two inputs.
Each of these two inputs can be connected to any of the four cell's inputs, and each of the four cell outputs can be connected to any other cell's input or the functional unit's output, as in
Fig.~\ref{fig:pe}.
The DFE matrix size is parametric
and can be customized at compile time to fit the target device resources. Programming the DFE means selecting all used inputs, outputs, and operators,
and routing all intermediate results.
Each node can serve as an operator, as a routing resource, or both.

\begin{figure}[t]
\lstset{  belowcaptionskip=1\baselineskip,
  language=C,
  showstringspaces=false,
  basicstyle=\footnotesize\ttfamily,
  keywordstyle=\bfseries\color{red},
  commentstyle=\itshape\color{purple!40!black},
  otherkeywords={1,2,3,5},
  keywords=[2]{1,2,3,5},
  keywordstyle={[2]\color{green!40!black}},
  identifierstyle=\bfseries\color{blue},
  stringstyle=\color{orange},}
\begin{lstlisting}[caption=Example code fragment with unavoidable branches, label=lst:for]
for (i = 0; i < M; i++) {
  for (j = 0; j < N; j++) {
    if (A[i][j] > B[i][j])
      C[i][j] = A[i][j]+3*B[i][j]+1;
    else
      C[i][j] = A[i][j]-5*B[i][j]-2;
  }
}
\end{lstlisting}
\end{figure}

\begin{figure}[t]
  \centering
  \includegraphics[width=.7\columnwidth]{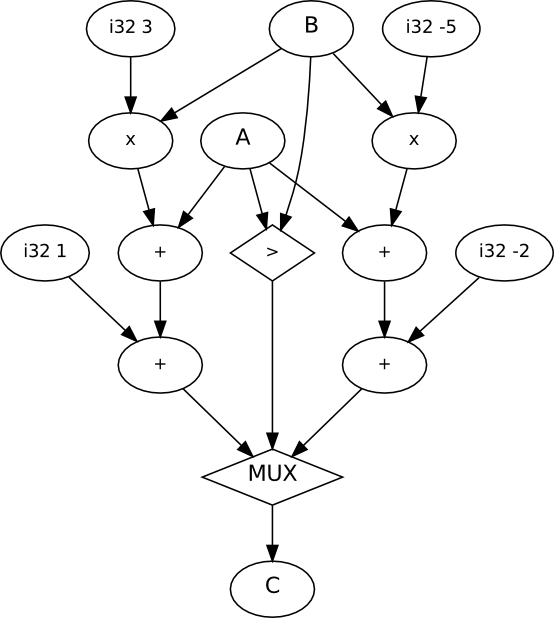}
  \caption{DFG extracted from Listing~\ref{lst:for}.}
  \label{fig:phi}
\end{figure}

To map source code onto the FPGA, we decided to extend the original version by adding comparison operators, as well as $MUX$ nodes enabling the implementation of
simple select statements directly in the FPGA fabric without requiring communication with the host processor. For instance, code with
dynamic branches (Listing 1) is converted in the DFG of Fig.~\ref{fig:phi} and can be directly executed on the DFE.
This allows us to operate on code blocks bigger than plain basic blocks, although we still cannot surpass loop's boundaries. As a consequence, the DFGs we manipulate are acyclic.
Another minor
improvement we made is to enable the transformation of inputs into constants. This is very simple to realize, as it requires only masking
one signal, but can considerably reduce the transfers needed from the host system.
Our implementation focuses on 32 bit operations and has some limitations: we do not support integer division nor remainder operations.
Only integer data types are currently supported.

\subsection{Place \& Route algorithm}
\label{ss:pr}
To place\&route the requested functional units in the DFE we have adopted a Las Vegas-type algorithm, that is, a stochastic algorithm that ends with a correct
solution --- if this solution exists.
The DFE, contrary to classical FPGAs, has no routing nodes and a Manhattan type topology; This makes the problem NP-complete~\cite{Vai00}.
This peculiarity prevented us from using off-the-shelf routing solutions, such as Verilog-To-Routing~\cite{Luu14}, and forced us to devise a custom, simple place-and-route algorithm.

Given the DFG of a piece of code, our algorithm works one node at a time.
It starts by randomly selecting a node, giving higher probabilities of being selected to input/output nodes (nodes directly connected to an input or output).
The number of interfaces on the border is very limited (equal to the perimeter of the overlay) thus we
favor the nodes having stricter positioning needs.
We then randomly select a candidate position, weighting the available positions using a narrow Gaussian distribution centered about the DFE's center.
This favors the positions closer to the border, leading to shorter paths.
This weight distribution, however, is altered to group nodes together, particularly so if two given nodes share an input or output (or both).
Having closer-by nodes has the additional advantage that free regions of the DFE can be easily reused to host other independent execution graphs.

Once a position is chosen, the algorithm finds the shortest paths for all node's inputs and outputs.
This is accomplished using Dijkstra's algorithm~\cite{Dijkstra59} from the node to all the DFE's cells where the desired variable is replicated, selecting then the closest option.
If the algorithm fails in finding any of the desired paths, the algorithm backtracks one step and randomly seeks a new position (excluding the failed one).
After a given number of failed attempts, the algorithm randomly selects a new node and starts again.
If the system runs out of nodes to place, it backtracks a random number of steps and starts from scratch from a previous setting.

Once a node is successfully routed, all previously-placed nodes are checked to see if either they provide an input to the current node, or if they take the node's output as input.
If this is the case, the connections are established.

Although there are many optimizations possible --- for instance, previously allocated nodes could be moved to reduce the paths' lengths ---, these strategies are strictly linked with the
chosen overlay and thus are beyond the scope of this paper.

\section{Results}
In this section we report three different performance analyses.
The first two individually address the main components of the system, namely the JIT framework and the DFE.
In the latter, we report a case study done on a full prototype in order to quantify the data management overhead.

\subsection{Software Framework}
To evaluate the performance of our software framework we selected
a set of standard benchmarks called PolyBench~\cite{polybench}. For each application in this set
we report the capability of the analysis step of detecting a potential opportunity
for off-loading, the compatibility with the implemented DFE, some statistics about
the generated DFG, and the time spent in the analysis. Results are summarized in Tab.~\ref{tab:polybench}.
As can be seen in Tab.~\ref{tab:polybench} we are able to detect almost all
SCoPs detected by Polly, and only in two cases our detection fails. In both codes
a problem managing $MUX$ nodes properly invalidates the analyzed SCoPs.
In two additional cases (\emph{nussinov} and \emph{floyd-warshall}), the system detects no SCoPs.
Unfortunately, our actual implementation for the DFE limits greatly the number of
applications that can be off-loaded for execution onto FPGAs, as divisions and floating point
operations are not supported yet. In one case ($heat-3d$), the extracted DFG is
very large, almost $300$ nodes, and our place\&route algorithm fails to map it on
the largest DFE we tested ($24 \times 18$).

\begin{table}
  \centering
  \caption{List of Polybench benchmark suite's algorithms for which our system was able to detect SCoPs (21/25),
    along with the corresponding DFG's statistics and the time needed for the analysis}
  \label{tab:polybench}
  \begin{tabular}{@{}lcccc@{}}
    \toprule
    {\bf Benchmark} & {\bf DFE off-load} & {\bf DFG nodes} &   {\bf Analysis} \\
                    &  &   {\bf in/out/calc. } & {\bf Time ($us$)} \\

 2mm  & Yes & 6/2/61 & 14209 \\
 3mm  & Yes & 9/3/85 & 28921 \\
 adi  & No, divisions & & 35249 \\
 atax & Yes & 6/2/49 & 8338 \\
 bicg & Yes & 6/2/49 & 7658 \\
fdtd-2d  & No, fp data &  & 33052 \\
gemm     &  Yes & 4/2/34 &		  7154                \\
gemver   & Yes &  13/4/95 &		 36500                \\
gesummv  & Yes &  8/3/70 &		 11723                \\
heat-3d  & Yes &  20/2/276	&	107645 \\
jacobi-1D & No, fp data &  &  7237 \\
jacobi-2D & No, fp data &  &   	 17757            \\
lu        & No, divisions  & &     	 18035          \\
ludcmp    & No, divisions  & &       37159 \\
mvt       & Yes & 6/2/40 &	  7028  \\
seidel    & No, divisions  & &         12296 \\
symm      & Yes &   6/2/64	&	 14659                \\
syr2k     & Yes &   6/2/52	&	  9112                \\
syrk      & Yes &   4/2/34	&	  5525 \\
trisolv   & No, divisions  & &   6646 \\
trmm      & Yes & 4/2/30 &  6540 \\
\end{tabular}
\end{table}

\subsection{DFE}
Before reporting the measurements done on the complete system we would like to
detail the performance of the employed DFE over several FPGA families.
One of our goals is indeed being able to support most existing FPGAs.
We report four cases in Tab.~\ref{tab:synth}, namely a low-end and a high-end
chip for the two main producers of FPGA, Xilinx and Altera.
For each device, we report resource utilization and maximum working frequency for different
DFE matrix sizes. The last line in each device summarizes the largest DFE that we were able to route. All reported data are for 32-bit wide data paths able to handle signed integer operations.

The required number of computing elements significantly varies from one FPGA to the other, as well as the maximum working frequency.
Routing our DFE is particularly critical once the size of the system exceeds $80\%$ of the available logic.
All results are obtained directly from the proprietary software of the two producers and no extra manual effort has been devoted to optimize the presented results. In~\cite{Capalija13} authors demonstrated how, with a better placement strategy, higher performances can be achieved: repeating the same results is out of the scope of this work.

It is interesting to note that even low-end FPGAs can be suitable for off-loading many of the algorithms presented in Tab.~\ref{tab:polybench}.

\begin{table*}
  \centering
  \caption{DFE resources' utilization on various devices}
  \label{tab:synth}
  \begin{tabular}{@{}llcrrrr@{}}
    \toprule
    {\bf FPGA Device} & {\bf Tool} & {\bf DFE Size} & $\mathbf{F_{max}}$ & {\bf Slice Reg (FF)} & {\bf LUTs} & {\bf DSP48} \\
    \midrule
    Spartan 6 &
    ISE &
    $3 \times 3$ &
    140 MHz &
    11521 (\ 6.3\%) &
    10968 (11.9\%) &
    9 (\ 5.0\%)\\
    (\emph{xc6slx150t-3fgg900}) &
    v.14.7&
    $6 \times 6$ &
    85 MHz &
    38340 (20.8\%) &
    36505 (39.6\%) &
    36 (20\%) \\
    &
    &
    $8 \times 8$ &
    68 MHz &
    65547 (35.6\%) &
    62451 (67.8\%) &
    64 (36\%) \\
    \midrule
    Virtex 7 &
    Vivado &
    $3 \times 3$ &
    240 MHz &
    11639 (\ 1.3\%) &
    9916  (\ 2.3\%) &
    9  (\ 0.3\%) \\
    (\emph{xc7vx690t-3ffg1761}) &
    v.2015.2.1 &
    $9 \times 9$ &
    192 MHz &
    83022 (\ 9.6\%) &
    70547 (16.3\%) &
    81 (\ 2.3\%) \\
    &
    &
    $15 \times 15$ &
    192 MHz &
    222298 (25.7\%) &
    187764 (43.3\%) &
    225 (\ 8.0\%) \\
    &
    &
    $24 \times 18$ &
    155 MHz &
    420981 (48.6\%) &
    353057 (81.5\%) &
    432 (12.0\%) \\
    \midrule
    Virtex 7 &
    Vivado &
    $18 \times 18$ &
    167 MHz &
    317517 (52.3\%) &
    265641 (87.5\%) &
    324 (11.6\%) \\
    (\emph{xc7vx485t-2ffg1761}) & v.2015.2.1 \\
    \toprule
    & & & & {\bf Registers} & {\bf ALMs} & {\bf MULT}$\mathbf{9 \times 9}$ \\
    \midrule
    Cyclone IV &
    Quartus II &
    $3 \times 3$ &
    120 MHz &
    7495 (\ 4.9\%) &
    12496 (\ 8.3\%) &
    18 (\ 2.5\%) \\
    (\emph{EP4CGX150DF31I7AD}) &
    v.13.1 &
    $6 \times 6$ &
    115 MHz &
    24740 (16.3\%) &
    43988 (29.4\%) &
    72 (10.0\%) \\
    &
    &
    $9 \times 9$ &
    106 MHz &
    52982 (34.8\%) &
    95670 (63.9\%) &
    162 (22.5\%) \\
    &
    &
    $10 \times 10$ &
    105 MHz &
    64839 (42.6\%) &
    117634 (78.5\%) &
    200 (27.8\%) \\
    \toprule
    & & & & & & {\bf DSP Block} \\
    \midrule
    Stratix V &
    Quartus II &
    $3 \times 3$ &
    250 MHz &
    7857 (\ 1.5\%) &
    6412 (\ 2.4\%) &
    9 (\ 0.5\%) \\
    (\emph{5SGSED8N2F45I2L}) &
    v.13.1 &
    $9 \times 9$ &
    232 MHz &
    56295 (10.7\%) &
    45992 (17.5\%) &
    81 (\ 4.1\%) \\
    &
    &
    $15 \times 15$ &
    220 MHz &
    150292 (28.6\%) &
    122805 (46.8\%) &
    225 (11.4\%) \\
    &
    &
    $24 \times 18$ &
    185 MHz &
    282304 (53.8\%) &
    209227 (79.7\%) &
    432 (22.0\%) \\
    \bottomrule
  \end{tabular}
\end{table*}

\subsection{Prototype}
\label{sec:proto}
In order to assess the behavior of the system in a fully functional prototype we installed
a Xilinx Virtex-7 FPGA VC707 Evaluation Kit on the PCIe slot of a
Intel core i7-4790 (3.60GHz)-based workstation with 32GB of RAM and Ubuntu 16.04 (kernel 4.4.0).
The third device in Tab.~\ref{tab:synth} reports the DFE implemented over the FPGA available on the VC707 Evaluation Kit.

\begin{figure}[t]
  \centering
  \includegraphics[width=\columnwidth]{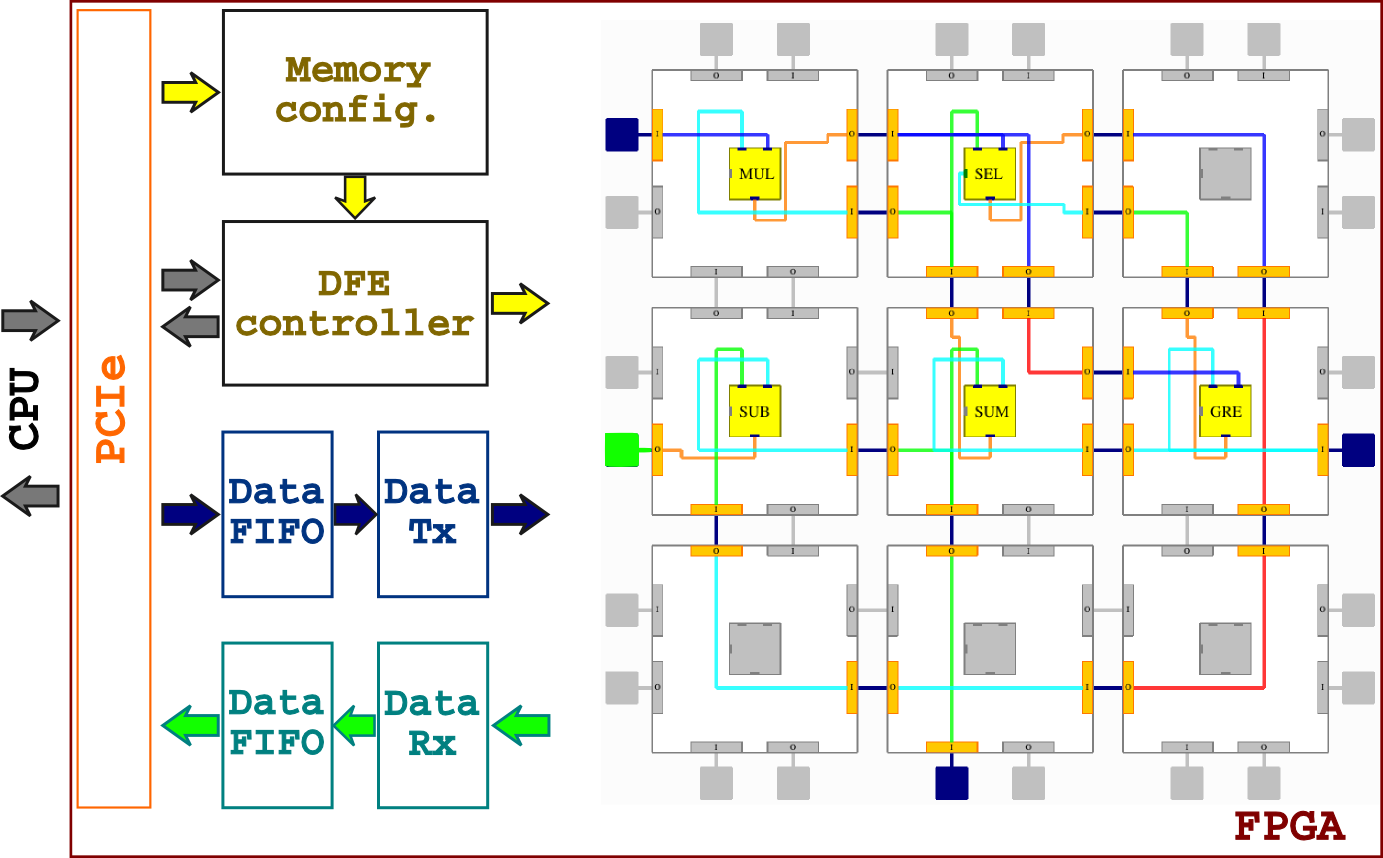}
  \caption{Block diagram of the prototype system highlighting CPU and DFE interfaces and control logic.}
  \label{fig:fpga_diag}
\end{figure}

\begin{figure*}[ht]
    \centering
    \begin{subfigure}[t]{.28\textwidth}
      \centering
      \includegraphics[height=2.5cm]{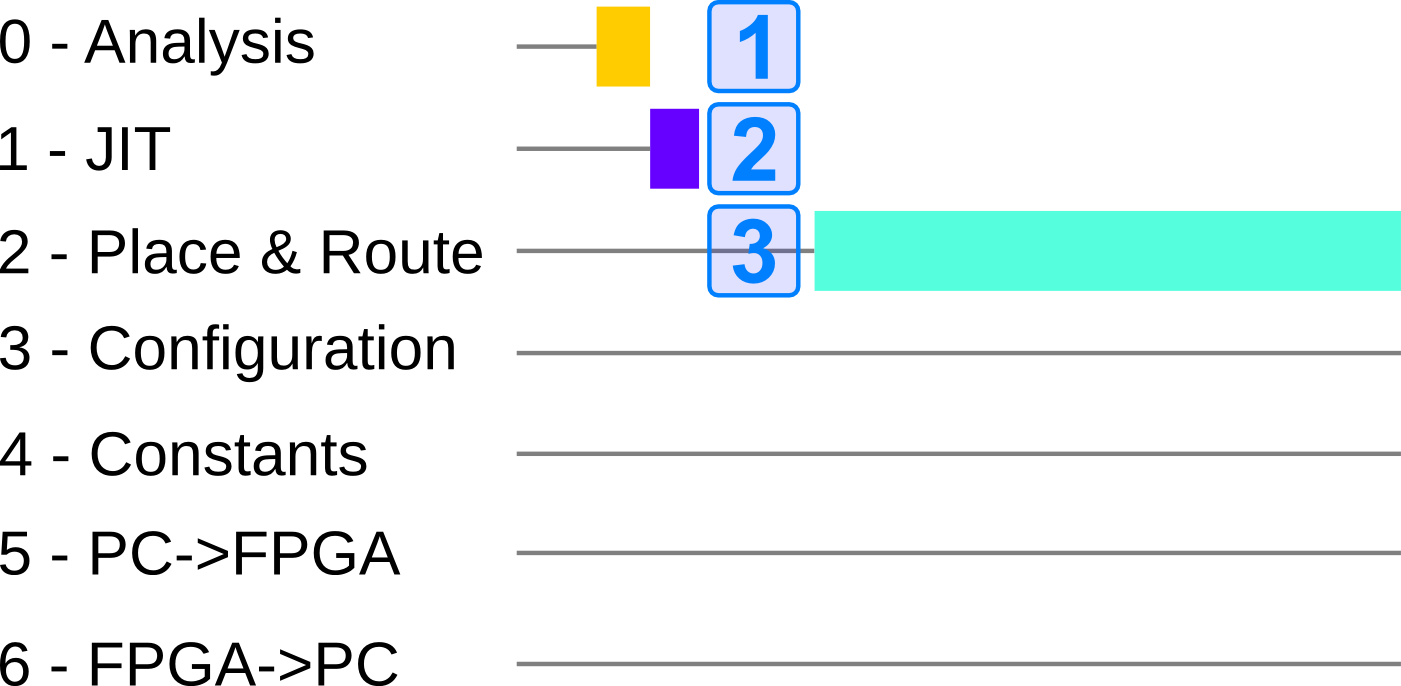}
      \caption{Analysis followed by JIT compilation and initial part of P\&R.}
    \end{subfigure}%
    \vspace{10pt}
    \begin{subfigure}[t]{.28\textwidth}
      \centering
      \includegraphics[height=2.5cm]{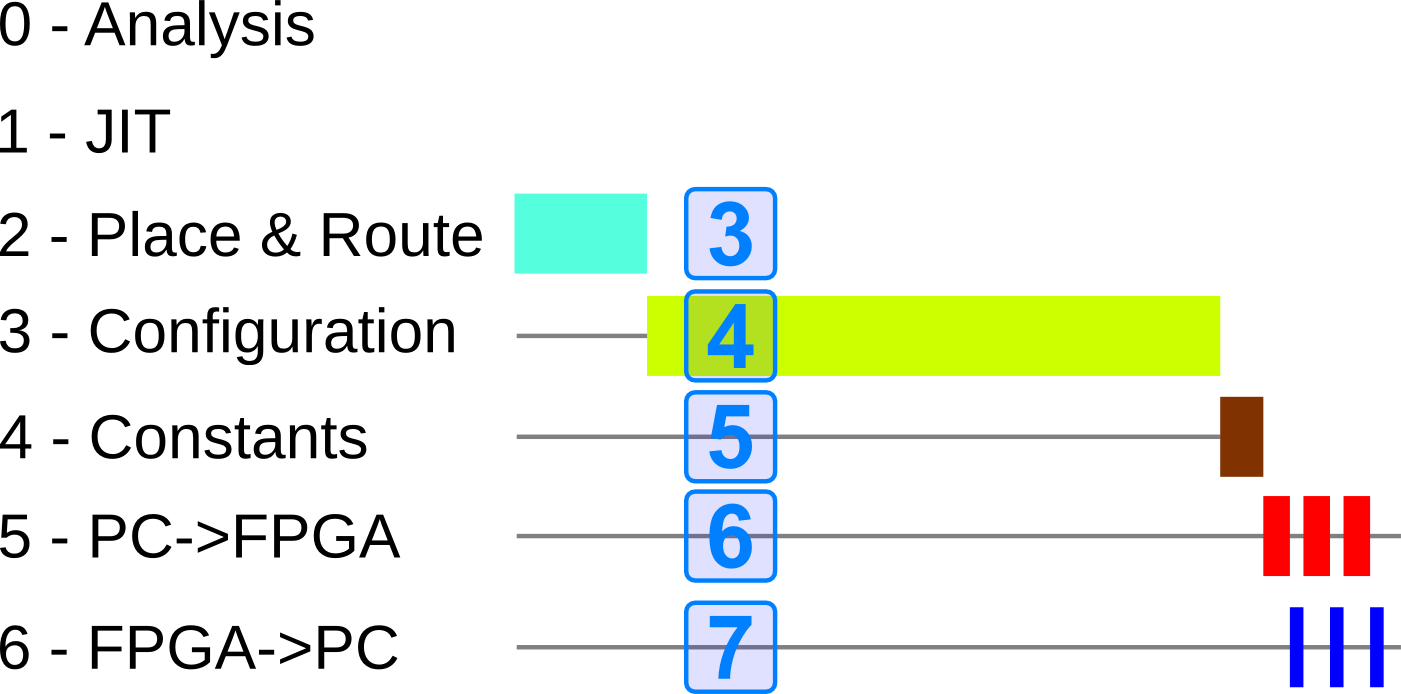}
      \caption{End of P\&R followed by DFE configuration.}
    \end{subfigure}%
    \vspace{10pt}
    \begin{subfigure}[t]{.28\textwidth}
      \centering
      \includegraphics[height=2.5cm]{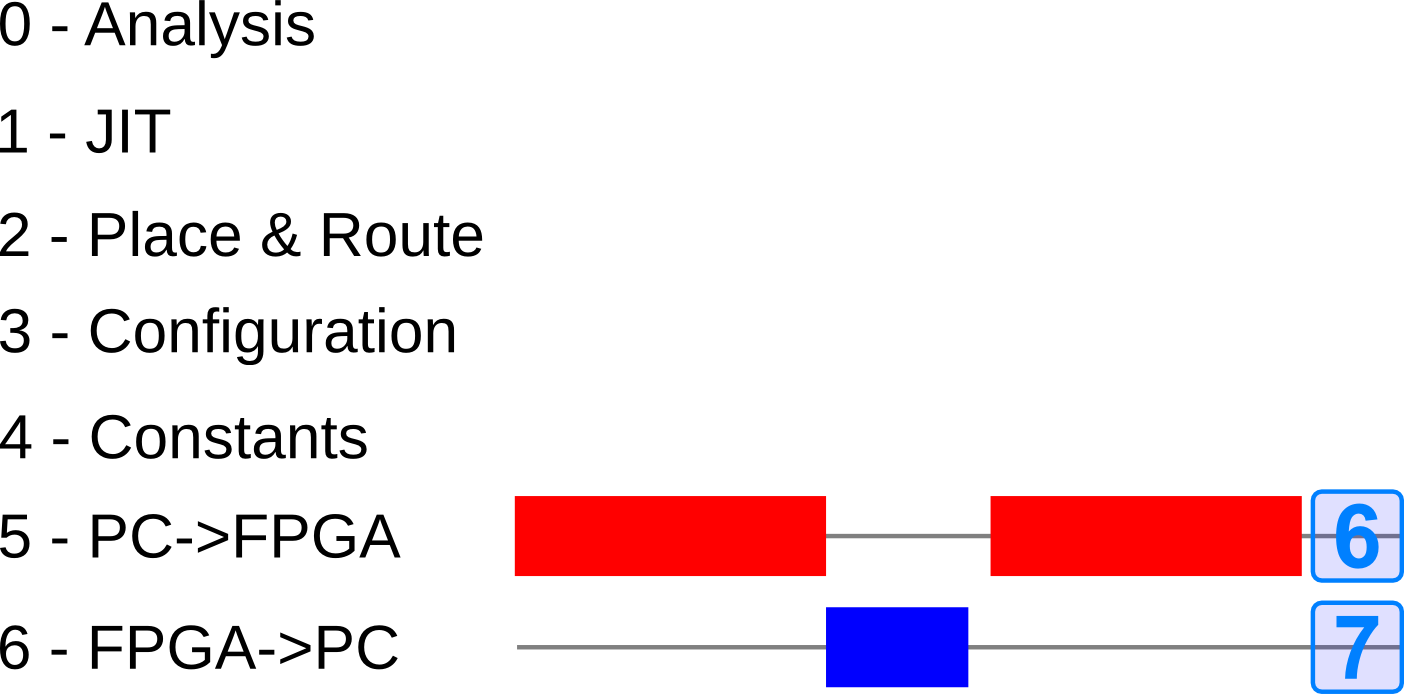}
      \caption{Data transfers to and from the DFE (detailed view).}
    \end{subfigure}%
    \caption{LTTng traces of all processing phases for the image processing example. Please refer to the text (Sec.~\ref{sec:proto}) for additional details.}
    \label{fig:traces}
\end{figure*}

To implement a full prototype, we added some glue logic around the PCIe core version $2.0$ equipped with $8$ lanes.  A block diagram of the realized system can be seen in Fig.~\ref{fig:fpga_diag}.
DFE's configuration switch and reset are implemented using a simple finite state machine controller.
Once the system is programmed, two additional basic
finite state machines handle data transfers to and from the host system. We decided to use in our implementation a very simple communication protocol between the host processor and DFE.
Each input and output data is augmented with a tag indicating the destination or the source respectively.
While this approach is inefficient in terms of bandwidth utilization, it is extremely flexible and significantly simplifies the
design of the I/O interface with respect to more efficient solutions.

In this scenario we decided to use a simple video processing application based on the OpenCV  video library in order to measure the overhead related to data management and transfer.
In this complete prototyping system, a video file is read from disk, processed with several convolution kernels and displayed on screen. After running the application for a few seconds,
the run-time decides to off-load the convolution function. At this stage, data are automatically transferred to and from the DFE through the PCIe interface and, if the requested data transfer
is above a programmable threshold, a DMA transfer is started.  Configurations and data are actually asynchronous and the system uses DMA transfers over the PCIe interconnect to both achieve
better transfer rates and to avoid requiring additional work from the CPU. The convolution selected for off-loading has a DFG with $17$ inputs, $1$ output, and $16$ computing nodes.

To precisely measure the time spent in the different processing phases, we augmented the application's and framework's code with LTTng events~\cite{lttng}.
In Fig.~\ref{fig:traces} we graphically represent the acquired events. Starting from Fig.~\ref{fig:traces}(a), the framework completes the analysis phase \myverb{1}
in $17.5ms$. This step assesses the off-loading opportunity, extracts the DFG and CFG of the code under evaluation, then immediately starts the JIT compilation \myverb{2} of the stub code.
JIT compilation is completed in $16.7ms$, then the DFG is passed to the place\&route routine \myverb{3}. The gap visible between these two phases is spent in the application outside our framework.
It is the time needed by OpenCV to read and decode some video frames. The same behavior can be periodically seen in the timing diagram. As stated before, the place\&route phase took a random time to
complete, in this example $1.18s$. Fig.~\ref{fig:traces}(b) depicts the end of the place\&route phase, followed by the downloading of the configuration \myverb{4} for the DFE.
It should be noted that, to highlight the last short phases in Fig.~\ref{fig:traces}(b), we used a zoom factor, so time scales of these images are different.
DFE configuration is completed in $2.1ms$. After that DFE is ready to work, the framework transfers all constants \myverb{5} (\SI{55}{\micro\second})
before sending data. As can be seen in Fig.~\ref{fig:traces}(c), data transfers are automatically broken in blocks and orderly transferred to the DFE.
DFE execution time is negligible, and soon after the PCIe bus is released by the write procedure, a read can start. Fig.~\ref{fig:traces}(c) stresses that the DFE is not continuously used.
This is related to the application code that manipulates results provided by the DFE before proceeding further and that PCIe is an arbitrated resource not always available.
Our PCIe data transfer protocol has a $75\%$ overhead, as we send $128$ bits for each $32$ bits --- for the sake of simplicity we perform no data compression while preparing
DMA packets.
With our simplified implementation we have measured data rates on the PCIe Gen2 $\times 8$ interface of approximately $230MB/s$,
and this figure has to be divided by 4 to account for the aforementioned lack of compression.
We can therefore expect to gain a significant speed-up by a sensible implementation of the transfer protocol --- for instance by integrating the RIFFA framework~\cite{Jacobsen15}, which
gets very close to the theoretical limit of 4GB/s --- but this goes beyond the scope of our paper.
Evidently we have a longer input transfer \myverb{6} to the DFE (\SI{35}{\micro\second}) compared to the
output \myverb{7} phase (\SI{16}{\micro\second}). The complete chain is able to process $31$ frames per second. Compared to the pure software implementation performance
--- almost $83$ frames per seconds --- this number is quite low, but this is a baseline proof of concept implementation that suffers from many limitations previously outlined.

\section{Conclusion}
In this paper we presented a fully transparent system able to
transfer code execution from a normal processor to an FPGA.
The proposed system is dynamic and able to adapt to changing
scenarios and system workloads. The approach
can be customized for specific applications, but it is also generic
and able to cope with a wide range of applications.
As demonstrated by extensive benchmarking our system can transparently
exploit FPGAs without requiring any developer skill or user interaction.
Performance is currently limited, but work is ongoing to address several limitations
and improve on them significantly.

\bibliographystyle{IEEEtran}
\bibliography{references.bib}

\end{document}